# Phason space analysis and structure modeling of 100 Å-scale dodecagonal quasicrystal in Mn-based alloy

Tsutomu Ishimasa[a*], Shuhei Iwami[a], Norihito Sakaguchi[a], Ryo Oota[a] and Marek Mihalkovič [b]

[a]*Graduate School of Engineering, Hokkaido University, Kita-ku, 060-8628 Sapporo, Japan;*
[b]*Institute of Physics, Slovak Academy of Sciences, 845 11 Blatislava, Slovakia.*

**Abstract**

The dodecagonal quasicrystal classified into the five-dimensional space group $P12_6/mmc$, recently discovered in a Mn-Cr-Ni-Si alloy, has been analyzed using atomic-resolution spherical aberration-corrected electron microscopy, *i.e.* high-angle annular dark-field scanning transmission electron microscopy (HAADF-STEM) and conventional transmission electron microscopy. By observing along the 12-fold axis, non-periodic tiling consisting of an equilateral triangle and a square has been revealed, of which common edge length is $a = 4.560$ Å. These tiles tend to form a network of dodecagons of which size is $(2+\sqrt{3})a \approx 17$ Å in diameter. The tiling was interpreted as an aggregate of 100 Å-scale oriented domains of high- and low-quality quasicrystals with small crystallites appearing at their boundaries. The quasicrystal domains exhibited a densely-filled circular acceptance region in the phason space. This is the first observation of the acceptance region in an actual dodecagonal quasicrystal.

Atomic structure model consistent with the electron microscopy images is a standard Frank-Kasper decoration of the triangle and square tiles, that can be inferred from the crystal structures of $Zr_4Al_3$ and $Cr_3Si$. Four kinds of layers located at $z = 0$, $\pm 1/4$ and $1/2$ are stacked periodically along the 12-fold axis, and the atoms at $z = 0$ and $1/2$ form hexagonal anti-prisms consistently with the $12_6$-screw axis. The validity of this structure model was examined by means of powder X-ray diffraction.

*Corresponding Author
 E-mail : ishimasa@eng.hokudai.ac.jp
 Telephone:   +81-11-706-6643         Fax:    +81-11-716-6175

1. **Introduction**

A quasicrystal is a solid or soft matter with bond-orientational order in a long range, exhibiting sharp diffraction spots with non-crystallographic symmetry such as 5-, 8-, 10-, 12-fold or icosahedral [1, 2]. From the geometric viewpoint, a quasicrystal is regarded as tiling of finite types of local structural units, or building bricks, of which edges are aligned along the directions satisfying the corresponding symmetry. Quasicrystals are usually classified into two categories by a way of arrangement of tiles; one is random tiling model and the other quasiperiodic model. In the framework of projection method [3 - 6], the quasiperiodicity is related to the periodicity in a higher dimensional space.

In the case of dodecagonal quasicrystal exhibiting 12-fold diffraction symmetry, the three-dimensional structure is understood as periodic stacking of thin layer consisting of two-dimensional quasiperiodic tilings, where the periodic axis is parallel to the 12-fold axis [5, 6]. Several types of the 12-fold quasiperiodic tiling in a plane [5 - 9] are known beside the random tiling [10 - 12]. They are composed of equilateral triangles and squares with the common edge length *a*. In some cases, the tiling also includes the third element that is either 30º rhombus or three-fold hexagon with 90º and 150º vertical angles (see for example Figure 4 in [13]). Such quasiperiodic tiling is constructed by hierarchic construction (inflation procedure) [7], application of matching rules [14], or projection of a periodic structure in the four-dimensional space [5, 6, 8, 9].

Actually, dodecagonal quasicrystals have been discovered in a wide range of materials from alloys to organic matters, namely Ni-Cr small particles synthesized by the gas-evaporation method [15], rapidly quenched V-Ni [16], Bi-Mn layer made by vacuum deposition [17], Ta-Te alloy [18], $BaTiO_3$ layer on Pt (111) surface [19], mesoporous silica [20], binary nanoparticles supperlattice [21], micelle-forming dendrimer [22] and ABC star polymer [23]. While this variety indicates that the dodecagonal order is universal in nature, the degree of structural perfection of them is low comparing with, for example, the stable icosahedral phase such as Al-Cu-Fe. Furthermore, they usually coexist with approximant crystals that is periodic tiling composed of the same local structural units as in the quasicrystal. This coexistence, or lack of single phase domain in other word, has prevented the detailed structural analysis of the quasicrystals. Recently new dodecagonal quasicrystal has been discovered in quaternary Mn-rich alloys containing Cr, Ni and Si [24]. Here, the five-dimensional space group was determined to be *P*12$_6$/*mmc* by analyzing the systematic extinction.

In this paper, we will focus on this dodecagonal quasicrystal. Based on the high resolution electron microscopy images, our aim is detailed structural analysis at the length-scales of several hundred Ångstroms. This involves quantitative characterization of both tiling arrangements and atomic decoration of each tile. Prior to reporting the experimental results, we review the necessary concepts described in the existence literature.

## 2. Dodecagonal tiling

Let us start from a triangle-square tiling in a plane with the common edge length $a$. Our tiling is either periodic or non-periodic. A vertex position $r_{//}$ of the tiling is expressed by the following formula using the four basis vectors $a_{j//}$ shown in Figure 1b;

$$r_{//} = \sum_{j=1}^{4} m_j a_{j//}, \quad |a_{j//}| = a. \quad \text{--- (1)}$$

According to the projection method, each vertex of tiling is assigned to a lattice point in the four-dimensional periodic lattice [5, 6], and then to the corresponding lattice point in the two-dimensional phason space, or perpendicular space, as expressed by the following;

$$r_{\perp} = \sum_{j=1}^{4} m_j a_{j\perp}, \quad |a_{j\perp}| = 1. \quad \text{--- (2)}$$

Here, $a_{j\perp}$ are vectors arranged as in Figure 1c. The choice of the length of these vectors is arbitrary, and it is assumed to be unity in this paper. The set of integers $m_j$ relates the points in two spaces; *i.e.* physical and phason spaces.

Figure 2a represents the ideal quasiperiodic tiling. In this tiling, one can find many dodecagons of which diameter is $(2 + \sqrt{3})a$. They are classified into two types according to the orientation of the inner hexagons. The two types are rotated by 30° with respect to each other. Their center positions are indicated by full (blue) or open circles (red) in Figure 2a. Linkages between these centers reveal another tiling of a bigger triangle and square, which are the second generation tiles enlarged by the factor of $2 + \sqrt{3}$. Self-similarity between the triangle-square tilings at these two length scales is a basis of Stampfli's inflation procedure [7] generating infinite "quasiperiodic" dodecagonal triangle-square tiling by recursively replacing tiling vertices with $(2 + \sqrt{3})$-deflated dodecagon clusters. In the resulting quasiperiodic tiling, all the end points of the vectors $r_{\perp}$ in the phason space fall within a strictly bounded domain named acceptance region: In the original version of the model, orientations of the dodecagons used in the inflation procedure are random, and introduce certain limited additional spread in the phason-space distribution. An alternative model based on the Stamfli's inflation construction uses a particular choice of strictly deterministic context-dependent cluster orientations [9]. The two models share the property of fractal-like phason space distribution [9, 25]. This apparently results from "dense disk-packing" constraint [26] leading to triangle-square tiling, as opposed to a general dodecagonal tiling including additional building tiles, for example 30º rhombus.

Besides these two tilings, random tiling [10 - 12] is also possible as shown in Figure 2b. Unlike above two models, fewer dodecagons appear in this model. In the quasiperiodic models, the vertex $3^6$ is always located at the center of the dodecagon, but not so in this model. (Hereafter, Schläfli symbol will be used to denote local configuration consisting of equilateral triangle and square.) There is no network of the centers of the dodecagons, and thus the higher generation tile appears only accidentally. One may also notice difference in local configurations. In the random tiling, there are many vertices of the type $3^3 \cdot 4^2$, which is shown by a double circle (green with black center), but not so in the deterministic model. The type $4^4$ appearing in the random tilling never occurs in the deterministic model. These differences are clearly demonstrated in the fourth and fifth columns in Table I. The phason-space distribution is generally broadened compared to the quasiperiodic models; furthermore, the broadening is increasing with increasing radius of the tiling patch selected in the physical space.

The distribution in the phason space is useful diagnostics to characterize the quasiperiodicity. In the case of periodic approximant, the end points of $r_\perp$ form a periodic lattice also in the phason space, if the phason strain tensor [27] is set to be zero. In this case, an elongated distribution appears in the phason space. Therefore, it is important to know the distribution of the end points of $r_\perp$ in order to characterize a tiling [20, 28]. In the case of 10-fold decagonal quasicrystal, the tiling observed by high resolution electron microscopy has been analyzed to characterize the quasiperiodicity [29]. Useful parameter $\eta$ for the evaluation of the quality of the tiling is defined by the following equation;

$$\eta = \sqrt{\frac{1}{n}\sum_{j=1}^{n}\left|r_{j\perp} - c_\perp\right|^2}. \qquad \text{--- (3)}$$

Here $n$ denotes number of vertices, and $c_\perp$ is the center position of the acceptance region. The factor $\eta^2$ is the variance of the displacement in the phason space. It is known that Bragg reflections appear if the factor $\eta$ is finite value independent to the number of vertices $n$ [28].

Numerical calculations for large models indicated that Stampfli's and the deterministic tilings have the same value of $\eta = 0.787$ independently of the sample size. Both types may exhibit Bragg peaks, and they are regarded as an ideal quasicrystal from the viewpoint of diffraction.

Beside the phason-space variance, we characterize observed dodecagonal tilings by several additional geometrical criteria, summarized below. As described below, the tiling observed experimentally includes very small number of 30º rhombi, hence this tile is omitted here.

(1) Number ratio of triangles and squares. This ratio $N_{tr}/N_{sq}$ is $4/\sqrt{3}$ for any dodecagonal quasicrystal, whether it is ideal tiling or maximally random one [10 - 12, 28].

(2) Number ratios of three types of the squares, and four types of the triangles, with different orientations. The former (latter) ratio is 1:1:1 (1:1:1:1) in the quasicrystal state.

(3) Frequency of the four types of the local configurations; namely $3^6$, $3^3 \cdot 4^2$, $3^2 \cdot 4 \cdot 3 \cdot 4$ and $4^4$.

(4) Frequency of two types of the dodecagons with inner hexagons differently oriented.

(5) Presence or absence of higher order generation tiles.

(6) Size and shape of the distribution in the phason space. The parameter $\eta$ may depend on the size of the distribution, and then its dependence on the physical space size is important. The value 0.787 for the ideal tiling is regarded as a reference. Asymmetry of the shape of the acceptance region is a sign breaking the 12-fold symmetry.

For the application of these criteria, the sample size must be large enough. From the experimental viewpoint, one needs atomic resolution in order to resolve each tile and, at the same time, wide sample area in order to examine the properties of the tiling.

## 3. Atomic decoration for structure modeling

In the framework of the projection method, a five-dimensional periodic crystal is usually assumed in order to analyze the three-dimensional structure of a dodecagonal quasicrystal. The following three space groups have been proposed for the five-dimensional crystal; $P12/mmm$, $P12_6/mmc$ ( or $P12_6/mcm$), and $P12/mcc$ [5, 6]. Bragg reflections are expressed using five indices;

$$\boldsymbol{g}_{//} = \sum_{n=1}^{5} h_n \boldsymbol{a}_n^* \qquad \text{--- (4)}$$

$$\boldsymbol{a}_n^* = \frac{1}{\sqrt{3}a}\left\{\boldsymbol{i}\cos\frac{(n-1)\pi}{6} + \boldsymbol{j}\sin\frac{(n-1)\pi}{6}\right\} \qquad \text{for } n = 1\sim 4,$$

$$a_5^* = \frac{1}{c} k .$$

Here three vectors *i*, *j* and *k* denote fundamental unit vectors orthogonal to each other. The vectors $a_1^*$ - $a_4^*$ are arranged as in Figure 1a, and span on the plane with the indices $h_1$ - $h_4$. The vector $a_5^*$ is parallel to the 12-fold axis. The parameter *c* denotes the period along the 12-fold axis. The space group with the *c*-glide or $12_6$-screw axis exhibits characteristic extinction in diffraction [5, 6].

In order to construct a realistic three-dimensional model, one needs to consider atomic decoration of triangle-square tiling. For this modeling, the σ-phase structure has been used as an approximant in the case of Frank-Kasper type [6, 15, 16, 30, 31]. The σ-phase is originally an intermetallic compound with the tetragonal unit cell, space group *P4₂/mnm*, which is categorized as tetrahedrally close-packed structure, *i.e.* Frank-Kasper type [31]. One can find $3^2 \cdot 4 \cdot 3 \cdot 4$ tesselation in the *c*-plane located at $z = \pm 1/4$. Well-known example is Fe-Cr σ-phase [32] with $a_\sigma$= 8.800 Å, $c_\sigma$= 4.544 Å . Here the common edge length *a* of triangle and square is 4.56 Å. This type of tiling with $3^2 \cdot 4 \cdot 3 \cdot 4$ configuration is found also in various kind of systems in which dodecagonal quasicrystals are formed [15, 16, 18, 20 - 23].

In the case of σ-phase structure, triangle- and square-bricks are decorated by atoms as shown in Figure 3b. In these bricks, atoms are located at four layers at $z = 0, \pm 1/4$ and 1/2. Their coordinates in the idealized models are summarized in Tables II and III. The triangle corresponds to a half unit cell of hexagonal $Zr_4Al_3$-type, and the square to a unit cell of cubic $Cr_3Si$- (A15) type structures. The period *c* along the *z*-axis is approximately equal to the common edge length *a* of the triangle-square tiling. The atoms at $z = \pm 1/4$ are located at the corners of the triangle and the square in the *c*-projection. This type of atomic decoration does not restrict connectivity of tiles as can be seen in Figure 3. Namely, any triangle-square tiling can be uniquely converted to a three-dimensional atomic model. The 30° rhombus shown in Figure 3e also appears as a defect in σ-phase (see Figure 5 in [33]).

Such a decoration model has been studied in detail in [6]. It has been shown that the five-dimensional space group of the decorated model is $P12_6/mmc$ that exhibits systematic extinction for the $h_1h_2h_2h_1h_5$ type reflections with odd $h_5$ index. This is just the extinction rule observed experimentally in Mn-Cr-Ni-Si quasicrystal [24].

4. **Experimental procedures**

The alloy specimen studied here is $Mn_{70.0}Cr_{7.5}Ni_{5.0}Si_{17.5}$ annealed at 700 °C for 230 h. As a starting material, pure chromium containing less than 100 ppm oxygen, Alfa Aesar 44005, was used. Other details of specimen preparation was described in [24]. In this alloy, the dodecagonal quasicrystal coexists with dominant β-Mn type phase as well as small amount of σ-phase. The specimen was crushed into fragments using an agate mortar and a pestle, and transferred on a microgrid mesh for the electron microscopy observation. Selected-area electron diffraction patterns were observed by a JEOL JEM200CS microscope at the acceleration voltage 200 kV. Thin specimens adequate for further observation were selected at this stage. Atomic resolution observation was carried out at 300 kV using a Titan$^3$ G2 60-300 microscope with double spherical aberration correctors installed at Laboratory of High-Voltage Electron Microscopy of Hokkaido University. Incident electron beam was set parallel to the 12-fold symmetry axis. For high-angle annular dark-field scanning transmission electron microscopy (HAADF-STEM), the convergence semi-angle of the probe was set to be 22 mrad, and the inner and outer acceptance semi-angles were set to be 63.8 and 200 mrad, respectively. The spherical aberration coefficient of the objective lens was adjusted to be -0.9 μm for the conventional high resolution imaging. In this case the specimen is illuminated by parallel beam, and the image is formed by interference between direct and diffracted waves. On the other hand, in the case of HAADF-STEM imaging, each atomic column is illuminated by a focused electron probe one by one. The atomic image can be regarded as atomic-number dependent contrast, Z-contrast, to a good approximation [34, 35].

The image contrast recorded in an area, 350 Å × 350 Å, was analyzed in order to examine the criteria (1)~ (6) described in Section 2. For this purpose, computer program was written that include four-dimensional indexing of image contrast using the formula (1). In order to make indexing easy, a weak feature in the image was eliminated beforehand by applying proper threshold intensity. Remaining weak spots located inside the tile was eliminated manually, if present.

## 5. Results and Discussions
### 5-1. Triangle-square tiling in real quasicrystal

Figure 4 presents a selected-area electron diffraction pattern exhibiting 12-fold symmetry. The size of the area was approximately 2500 Å in diameter. The reflections observed in this pattern can be indexed using the formula (4) with $h_5 = 0$. This observation implies presence of 12-fold bond-orientational order. The region corresponding to this diffraction pattern was further studied by means of electron microscopy.

An example of HAADF-STEM images is presented in Figure 5. In this image, one may notice bright spots forming triangle-square tiling with the edge length 4.6 Å. In Figure 5, the spots located at the corners of the tiles are brighter than other spots appearing inside the tiles. This is due to the fact that the atoms located at z = 1/4 and -1/4 are superimposed along the projection direction (see Figure 3). All squares have similar internal distribution of intensity looking like a "cross", and the triangles have the internal image like a "tripod". These characteristics agree perfectly with the atomic positions in the models shown in Figure 3. This agreement supports the structure modeling based on the $Cr_3Si$- and $Zr_4Al_3$-type structures, which will be further examined below.

There are three types of the local configurations, *i.e.* $3^2 \cdot 4 \cdot 3 \cdot 4$, $3^3 \cdot 4^2$ and $3^6$, corresponding to Figures 3b - d, respectively. The second is rather rare; an example is depicted in Figure 5. Each edge of these tiles is parallel to one of the twelve directions turning by 30°. Then, there is the 12-fold symmetric bond-orientational order in this structure. However, at a glance, there seems no translational regularity.

At each vertex of the triangle-square tiling there is a ring image with the diameter of 4.6 Å. This is the distorted dodecagon corresponding to the projection of the hexagonal anti-prism in Figure 3, of which size is approximately 4.6 Å (= *a*) in diameter. There is also bigger dodecagon as depicted in the upper left in Figure 5. This dodecagon can be regarded as the second generation one enlarged by the factor of 2 + √3. Its diameter is (2 + √3)*a* ≈ 17.0 Å. By connecting the centers of such dodecagons, it is possible to see the second generation triangle and square, while in some places they form a rectangular tile corresponding to a unit cell of K-phase [36]. Remarkable feature of the observed tiling is that any center of $3^6$ configuration is always the center of 17 Å dodecagon. This feature is different from the random tiling model in Figure 2b, but similar to the quasiperiodic model in Figure 2a.

In Figure 5, there are two 30° rhombi with the edge length *a*. Occurrence of the rhombus was examined in eight similar images taken at different places. Only five rhombi were detected in the area 5.9×10$^4$ Å$^2$ in total, hence their density may be the order of 10$^{-4}$ Å$^{-2}$. This result suggests that the 30° rhombus can be regarded as a defect in this specimen. The 30° rhombus is always isolated, and no aggregate of rhombi as in Figure 2 in [13] was observed.

The second atomic resolution image is presented in Figure 6, which is the interference electron microscopy image. Here, the 17 Å dodecagons are highlighted. It is noted that most atoms in this region participate to form a dodecagon, and the rest forms glue tiles also consisting of triangles and squares with 4.6 Å edge. There are two kinds of linkages of the dodecagons. One is edge sharing, and the other is overlapping. The former has $(2 + \sqrt{3})a \approx$ 17.0 Å separation, and the latter $(1 + \sqrt{3})a \approx 12.5$ Å. The long linkage forms either the second generation square or triangle tiles. The combination of two linkages forms the K-phase rectangle. The short linkage forms a smaller triangle corresponding to a half unit cell of F-phase [36]. However, the short linkage does not occur in either the Stampfli's, or the deterministic model. The F-phase triangle is rare in this image, and can be found near the lower margin in Figure 6. As indicated by thin lines, further bigger dodecagon tends to be formed, which is the third generation dodecagon enlarged by the factor of $(2 + \sqrt{3})^2$, but only a part is completed.

The internal distribution of intensity in each tile seems consistent to the structure models presented in Figure 3, but the coincidence is not as apparent as in the case of HAADF-SETM image described above. Interpretation of image contrast of such a non-periodic structure is an important problem to be studied in the future.

In Figure 7a, the third image is presented that covers relatively wide area, approximately 350 Å × 350 Å. The Fourier-spectrum of the image in Figure 7b again shows the 12-fold symmetry. The contrast of the image in Figure 7a is high, and the shapes of tiles are so exact that tiling could be extracted using the indexing formula (1). The result is presented in Figure 8. Using this tiling model, the geometric characteristics are examined. We will use a Cartesian coordinates system on Ångstrom scale hereafter referring to the position of the tiling vertices.

Numbers of tiles and local configurations counted in Figure 8 are summarized in the second column in Table I. Here, 5093 vertices and 7540 pieces of tiles were recognized. The number ratio of triangles and squares, $N_{tr}/N_{sq}$, is equal to 2.331, while the ideal one is $4/\sqrt{3}$ =2.309. Considering the finite size effect, they are regarded as equal. Number ratio of four types of the triangles with different orientations is 1238 : 1241 : 1397 : 1397 = 1 : 1.002 : 1.128 : 1.128. The ratio of three types of the squares is 816 : 710 : 736 = 1 : 0.870 : 0.902. These ratios are approximately equal to unity.

As a local configuration, $3^2 \cdot 4 \cdot 3 \cdot 4$ type appears most frequently, and $3^6$ type is also common. The $3^3 \cdot 4^2$ type appears, but less frequently comparing with both in the deterministic and the random tiling models. The $4^4$ type is almost absent. Unique example can be found at position $x = 230$ and $y = 75$. Besides the triangle and square tiles, there are five 30° rhombi. An example is found at the position $x = 125$ and $y = 195$. A pair of open (green) circles indicates this type of rhombus.

The center positions of the 17 Å dodecagons are shown in Figure 9 using full (blue) and open (red) circles. These two types will be referred as "Blue-type" and "Red-type" hereafter. In Figure 9, the Blue-type appears often rather than Red-type, and the number ratio is 208 : 127=1 : 0.61. Two types of linkages, $(2 + \sqrt{3})a$ and $(1 + \sqrt{3})a$ are drawn in this figure. One may notice the existence of a "hole" of network at which the 17 Å dodecagon is absent. Such a hole is also composed of mainly $3^2\cdot 4\cdot 3\cdot 4$ type with a few $3^3\cdot 4^2$ and $3^6$ vertices.

The tiling model in Figure 9 looks like complicated aggregate of several kinds of small crystallites. Then, as the first step, we will examine this tiling from the viewpoint of periodic crystal. However, as described in the next section, there are 100 Å-scale quasicrystal domains hidden in this tiling.

In the upper left in Figure 9 ranging between $0 < x < 190$ and $160 < y < 345$, there is a relatively dense network with the linkage $(2 + \sqrt{3})a$. One can notice presence of small domains of the second generation triangle. An example is seen at $x = 60$ and $y = 250$. This region corresponds to the hexagonal approximant observed in Mn-(V, Cr)-Si systems [37, 38]. Furthermore, there is also a domain of the second generation square, for example, at $x = 40$ and $y = 290$.

In the upper right, there are two "fan-like" domains. One domain ranging between $170 < x < 240$ and $160 < y < 345$ consists of Red-type dodecagons, and the other between $220 < x < 350$ and $140 < y < 330$ of Blue-type. In each domain, tiling has one-dimensional periodicity with $(1 + \sqrt{3})a$ in a limited range, but has no periodicity in the perpendicular direction. In the former Red-type domain, narrow bands of σ- and K-phases seem to occur, and in the latter Blue-type domain three types, namely σ-, K- and F-phases [36] appear. Then, as a whole, they look like neither crystal nor dodecagonal quasicrystal. In the lower right region between $200 < x < 355$ and $0 < y < 140$, the network of $(2 + \sqrt{3})a$ is not complete, and the second generation tiles are not common. This area is again not a periodic crystal.

It is noted that there is a topological defect named disclination at the lower right corner at $x = 335$ and $y = 40$, which looks like a crack in Figure 9. Near the disclination, distorted tiles appear in the experimental image in Figure 7a. Except for this, the bond-orientational order in the 12 directions is strictly fulfilled in the whole region. This analysis of the network with the two types of the linkages seems to indicate complicated coexistence of small crystallites as well as absence of periodicity in a long-range.

### 5-2. Phason space analysis

As described in Section 2, the analysis in the phason space effectively reveals long-range translational order related to the quasiperiodicity. For this purpose, the phason components in the observed tiling were analyzed using the formula (2). The results are presented in Figure 10 using the same color scale both in the phason and physical spaces. This analysis indicates that there are three isolated domains in the phason space, and that there are also corresponding three domains in the physical space. Each domain in the phason space seems to have a circular shape with the radius smaller than $2|a_{j\perp}| = 2$. These three domains have different centers.

The upper left area in Figure 10c was extracted for further analysis. This area corresponds to the central domain in Figure 10a, and is presented in Figure 11a. In order to examine the development of the distributions, eight concentric domains were selected in the physical space. They include 272 - 1405 vertices as depicted by thick lines in Figure 11a. Their diameters range from 66 to 181 Å. The corresponding distributions are shown in Figures 11c - j. In Figures 11c - g, the distributions look like a circle with almost constant diameter, while the shapes are slightly elongated in the largest three cases shown in Figures 11h - j. When the number of vertices is increased step by step, the vertices newly appearing fill the fixed domain, namely the acceptance region, to be more dense in particular in Figures 11c - g. One can even notice the generation of self-similar like pattern in the phason space by comparing distributions in Figures 11c and 11d ( and also Figures 11d and 11e).

Figure 12 presents the relation between the factor $\eta$ and the diameter of the concentric region. Up to the diameter approximately 120 Å, the factor $\eta$ is nearly constant between 0.82 and 0.88, and increases in rather steep slope from the diameter 130 Å. This increase corresponds to the elongation of the distributions in Figures 11h - j. The value, $\eta = 0.88$, is compared with those in the ideal dodecagonal tilings, 0.787, and also in the random tiling, 1.072, listed in Table I. The experimental value of $\eta$ is located in between them. These phason space analyses, shown in Figures 11 and 12, clearly indicate that the structure is not the random tiling, but holds the property very similar to the "quasiperiodicity" up to the diameter of about 130 Å, while the structure is not exactly the same as the Stampfli's or the deterministic model. It is noted that the structure would exhibit Bragg reflections, if the equality of the value $\eta$ was maintained in infinitely large region. Accordingly the region satisfying the near-equality of $\eta$ is regarded as a high-quality quasicrystal.

The factor $\eta$ was calculated also for other two regions, *i.e.* upper (red) and lower (green) regions in Figure 10a. The domain presented in Figure 13d was selected from the lower right in Figure 9 ranging between $240 < x < 350$ and $30 < y < 170$. This domain includes 750 vertices. The corresponding distribution is a circle as seen in Figure 13b. This is slightly smeared comparing with the former ones presented in Figure 11, and the factor $\eta$ was calculated to be 1.029. The third domain ranges between $170 < x < 310$ and $200 < y < 340$, and is depicted in Figure 13e. In this case, the distribution shown in Figure 13c is slightly elongated. The factor $\eta=1.179$ was calculated for 759 vertices. This domain is located at the boundary of Red- and Blue-type fans described above. The increase of size in the physical space causes the participation of one-dimensional periodic structures existing in both sides, and thus much scattered distribution in the phason space.

An intriguing result of this study is the contradictory pictures with respect to the same structure; One is the complicated aggregate of small crystallites described in the previous section, and the other is the finite domains of quasicrystal. This is due to the fact that the present state is intermediate between two extremes, *i.e.* the perfect periodic crystal and the perfect quasicrystal. As revealed by the phason space analysis, the quasicrystal picture has advantage. However, one cannot cover the whole region only by quasicrystal domains with well-defined circular acceptance regions, but with slightly elongated ones corresponding to the local periodicity in part. Accordingly we conclude that the present state consists of 100 Å-scale domains of high and low-quality quasicrystals plus small crystallites located at their boundaries.

### 5-3.  Atomic structure model : evidence by powder diffraction

In this section, an atomic structure model of the dodecagonal quasicrystal is proposed. This modeling was carried out by decorating the observed triangle-square tiling by atoms as in the structure models show in Figure 3. The common edge length $a$ of the tiling and the period $c$ along the 12-fold axis were assumed to be 4.560 Å and 4.626 Å, respectively [24]. Here we assumed two kinds of atoms, *i.e.* a metallic atom "M" and Si. This is due to close similarity of three kinds of metal, namely Mn, Cr, Ni, with respect to the atomic scattering factors for both electron and X-ray. The alloy composition was assumed to be $M_{85}Si_{15}$ referring to the experimental one $Mn_{74}Cr_{10}Ni_1Si_{15}$ [24]. By ignoring the small amount of Ni, we assumed that $Mn_{0.88}Cr_{0.12}$ corresponds to M. With respect to the position of Si atom, the structure model of the $Mn_{70}Cr_{10}Si_{20}$ σ-phase was referred [39]. In the present model, Si atoms occupy CN12 sites with the occupation probability 0.44 Si and 0.56 M as shown in Tables II and III. The term "M/Si" denotes this type of atom.

Figure 14 presents the structure model including 3584 M and 1849 M/Si atoms, of which composition is $Mn_{75}Cr_{10}Si_{15}$. The size of this model is 132 Å in diameter and corresponds to the fourth largest circle shown in Figure 11a. Details of the model is summarized in the third column in Table I. One 30° rhombus is included in this model. The atomic decoration of the rhombus has not been solved experimentally, but can be deduced from the boundary condition of the surrounding triangles and squares[6, 13, 33]. The atomic coordination near the 30° rhombus is not Frank-Kasper type, but with coordination number CN13 (for details see [13].) If one look Figure 14 at lower grazing angle, it is noticed that the quasi-lattice lines penetrate the model entirely along the directions indicated by six arrows.

Now it is possible to calculate X-ray diffraction intensity of this model. Here we assumed Cu K$\alpha_1$-radiation. Each reciprocal plane was divided into 2000 × 2000 parts in every 0.001 Å$^{-1}$ for the calculation. Three calculated patterns at $h_5 = 0$, 1 and 2 are presented in Figure 15. In these patterns, the reflections are arranged in 12-fold symmetric manner, which are indexed by the formula (4). Strong reflections accompany a weak subsidiary maximum that reflects the sample size, 132 Å in diameter. The size-limited shape of the reflections is another evidence of the highly ordered quasicrystal.

At the plane $h_5 = 4$ ($h_5 = 3$), similar pattern to that at $h_5 = 0$ ($h_5 = 1$) with slightly weaker intensity was calculated. The similarity between $h_5 = 0$ and 4 is due to the four-layer stacking structure located at exactly $z = 0$, ±1/4 and 1/2. The other similarity between $h_5 = 1$ and 3 is due to the same feature plus presence of a mirror plane at $z = 0$. In Figure 15, the reflections of $h_1h_2h_2h_1h_5$ type with odd $h_5$ index are not visible. An example is extinction of 1 2 2 1 1 in Figure 15b, while the 1 2 2 1 2 reflection in Figure 15c has rather strong intensity. This is due to the presence of $12_6$-screw axis, and $c$-glide, in the corresponding five-dimensional structure, and is consistent to the atomic decoration scheme in Figure 3 [6]. It is noted that the difference in intensity distributions in Figures 4 and 15a is due to multiple scattering effect in the electron diffraction [6].

In the next step, the integrated intensities of reflections were estimated in the reciprocal planes. A powder X-ray diffraction pattern was calculated using the integrated intensity by assuming a spherical shape with a diameter 132 Å, and also the atomic displacement $B$-factor 2.0 Å$^2$. In this calculation the contribution of K$\alpha_2$-radiation was also included, and pseudo-Voigt function with mixing parameter 0.5 was assumed for each peak. In Figure 16, the calculated pattern is compared with the experimental one [24]. While the experimental pattern exhibits some artifacts due to the subtraction process of coexisting β-Mn type phase, overall agreement is good. This agreement implies the following;

(1) The present structure model based on the triangle-square tiling with the atomic decoration is essentially correct to describe the structure of the quasicrystal.

(2) The contribution of the small crystallites is not clearly visible possibly because of their small amount, and also because of their structural similarity to the quasicrystal.

It is added that the measured diffraction pattern has slightly sharper peaks than the calculated one with respect to the following reflections; 0 0 0 0 2, 1 1 0 0 2, 2 3 1 $\bar{1}$ 2, 2 2 0 $\bar{1}$ 3 and 0 0 0 0 4, all having relatively large $h_5$ index. This may be due to the needle-like shape of the quasicrystal elongated possibly along the 12-fold axis (see Figure 4 in [24]).

## 5-4.  Toward highly ordered dodecagonal quasicrystal

The two basic structure models, *i.e.* the quasiperiodic model and the random tiling model, are different with geometric characteristics, and also different with the stabilization mechanism as a realistic phase. In the former model, the quasicrystal has the lowest internal energy as a ground state, but in the latter it is expected that the quasicrystal is stabilized by the configurational entropy [10-12, 39] with respect to the tile rearrangement at high temperature. As described above, the observed tiling is different from the random tiling model, but shows similarity to the quasiperiodic model.

In reality, our system is not the two-dimensional, but the three-dimensional solid. "Stacking disorder" might occur along the 12-fold axis, if the columns perpendicular to the dodecagonal plane could be interrupted. In relation to such stacking disorder, the role of the 30º rhombus has been discussed [13, 33]. However, in the present case, the density of the rhombus is very low in the order of $10^{-4}$ Å$^{-2}$. Furthermore, the electron diffraction experiment has revealed no diffuse streak along the 12-fold axis (see Figures 1b and c in [24]). The HAADF-STEM image contrast, related to each atomic column, exhibited no indication breaking of the periodicity along the 12-fold axis. Accordingly, the two-dimensional picture may be a good approximation in the present case, and the configurational entropy may not play an important role to stabilize the quasicrystal.

From these considerations, one may expect an energetically stabilized dodecagonal quasicrystal in the Mn-based alloy. However, the problem seems not so simple as follows;

(1) Both the Stampfli's model and the deterministic model include only one type of linkage, $(2 + \sqrt{3})a$, between the 17 Å dodecagons. On the other hand, the tiling actually observed have two types of linkages, *i.e.* $(1 + \sqrt{3})a$ and $(2 + \sqrt{3})a$, and shows relatively small phason-space variance in spite of the occurrence of the shorter one. This suggests possibility of different type of quasiperiodic model including both types of linkages, although no such geometric model has been proposed.

(2) In the ternary Mn-Cr-Si system, the σ–phase forms as a high-temperature phase, and the hexagonal phase corresponding to the second-generation triangle as a low-temperature phase [38]. The K-phase rectangle was also observed at boundary of the domains of the hexagonal phase as seen in Figure 10 in [38]. However, no dodecagonal quasicrystal has been observed in the Mn-Cr-Si alloy. The addition of a few atomic percents of Ni into the ternary system promoted the formation of the quasicrystal. At the tiling level, number of $3^3 4^2$ type vertex was increased. This fact suggests that Ni atom occupies specific sites in the triangle and square tiles, and further plays important role to form the dodecagonal quasicrystal. This situation recalls to us the possibility of matching rule or weak matching rule [40] in a dodecagonal quasiperiodic tiling. However, at present, only one type of matching rule is known for the dodecagonal tiling that includes the third type of tile [14].

(3) The distribution of the 17 Å dodecagon is not uniform in the present specimen as indicated by the existence of the "hole" in the network. This non-homogeneity suggests that the present structure is not in equilibrium but in some frozen state. In order to clarify the equilibrium state at the low temperature and also to reach a highly ordered quasicrystal, further annealing experiments are certainly necessary.

## 6. Conclusion

The dodecagonal quasicrystal formed in the $Mn_{70.0}Cr_{7.5}Ni_{5.0}Si_{17.5}$ alloy was investigated at the atomic scale. Non-periodic tiling was observed, in which the equilateral triangles and the squares with common edge length 4.560 Å are arranged. This tiling was further interpreted as an aggregate of 100 Å-scale domains of high- and low-quality quasicrystals with small crystallites located at their boundaries. In the whole region, the 12-fold symmetric bond-orientational order is well maintained. A network of the 17 Å dodecagons was observed, which frequently form the second generation tiles, *i.e.* bigger equilateral triangle and square scaled by the factor of $2+\sqrt{3}$. Furthermore, a part of the third generation dodecagon was also observed. Corresponding to these results, a densely-filled circular acceptance region was recognized in the phason space for the first time as a dodecagonal quasicrystal. These features indicate that the present structure is not explained as the random tiling model. The elementary triangle and square tiles are decorated by atoms according to the Frank-Kasper type close packing. This decorated model explained well the intensity distribution in the powder X-ray diffraction pattern.

In conclusion, the Mn-Cr-Ni-Si quasicrystal is located near to the perfect dodecagonal quasicrystal, and can be the good starting point of the future study of this type of quasicrystal.

**Acknowledgements**

The authors thank Hayato Iga and Yuya Tanaka for their cooperation to this research project. They also thank Marc de Boissieu for valuable discussions. A part of this work was supported by "Nanotechnology Platform" Program of the Ministry of Education, Culture, Sports, Science and Technology (MEXT), Japan. MM was supported by Slovak grants VEGA 2/0189/14 and APVV-0076-11.

**Figure Captions**

Figure 1

Basis vectors of dodecagonal quasicrystal. (a) in reciprocal physical space, (b) in direct physical space, (c) in direct phason space after [5].

Figure 2

Triangle-square tiling. (a) deterministic model and (b) random tiling model. Vertices are classified into four types according to their local configuration. Vertices of type $3^2 \cdot 4 \cdot 3 \cdot 4$ and $3^3 \cdot 4^2$ are indicated by small full circle (black) and small double circle (green with black center), respectively. Vertex $4^4$ indicated by full circle with spot (red with black center) appears only in (b). Vertices of $3^6$ are denoted by larger open circle (red) or larger full circle (blue). (b) was redrawn with some modification from Figure 1 in [11].

Figure 3

Four local configurations and 30° rhombus with atomic decoration. (a) $4^4$ corresponding to $Cr_3Si$-type structure, (b) $3^2 \cdot 4 \cdot 3 \cdot 4$ to σ-phase, (c) $3^3 \cdot 4^2$, (d) $3^6$ to $Zr_4Al_3$-type, and (e) 30° rhombus . Darker (blue) and brighter (green) circles denote "M" and "M/Si" atoms, respectively. Open, full, and double circles correspond to atoms located at z = 0, 1/2 and ±1/4, respectively. For the assignment of sites A - J, see Tables II and III.

Figure 4

Selected-area electron diffraction pattern of the dodecagonal quasicrystal formed in the $Mn_{70.0}Cr_{7.5}Ni_{5.0}Si_{17.5}$ alloy. The indices of reflections are A: 1 2 1 0 0, B: 2 3 2 0 0, C: 1 2 2 1 0 and D: 1 3 3 1 0.

Figure 5

HAADF-STEM image. Notice triangle-square tiling with the common edge length 4.6 Å. The larger triangle and square are the second generation tiles with 17.0 Å edge.

Figure 6

Conventional high resolution electron microscopy image observed along 12-fold axis. Diameter of the dodecagon is 17.0 Å. The factor $\eta = 0.943$ was calculated for the triangle-square tiling observed in this image.

Figure 7

(a) Conventional high resolution electron microscopy image observed along 12-fold axis and (b) its Fourier transform. Letter A in (b) indicates 1 3 3 1 0 reflection.

Figure 8

Triangle-square tiling extracted from high resolution image presented in Figure 7a. Vertices are classified according to their local configurations as in Figure 2. Four basis vectors $a_{1//}$ ~ $a_{4//}$ are inserted that are used for indexing. Notice crack in the right lower corner that corresponds to a disclination.

Figure 9

Network of dodecagonal centers with two kinds of linkage, $(1 + \sqrt{3})a$ and $(2 + \sqrt{3})a$. Two domains surrounded by (orange) lines correspond to two quasicrystal domains shown in Figure 13, which were used for analysis of the factor $\eta$.

Figure 10

(a) Distribution of 5031 vertices in the phason space. A portion near the disclination was omitted in this analysis. (b) Basis vectors $e_{1\perp}$~$e_{4\perp}$ used for the construction in the phason space. Their length is regarded as standard. (c) Distribution of 5031 vertices in the physical space with the same color scale as in (a). Each colored domain corresponds to quasicrystal regions with different centers in the phason space.

Figure 11

(a) Triangle-square tiling selected from Figure 10c ranging between 0 < x <190 and 160 < y < 345. Concentric domains drawn by thick lines were used for the phason-space analysis shown in (c) - (j). (b) Basis vectors $e_{1\perp}$- $e_{4\perp}$ used for the calculation in the phason space. (c) Distribution for the smallest domain of 66 Å in diameter with 202 vertices. (d) 82 Å with 307 vertices, (e) 99 Å with 437 vertices, (f) 115 Å with 586 vertices, (g) 132 Å with 760 vertices, (h) 148 Å with 953 vertices, (i) 165 Å with 1170 vertices, and (j) 181 Å with 1405 vertices. Notice appearance of self-similar patterns in (c) - (f).

Figure 12

Dependence of the factor $\eta$ on the diameter of the concentric domains. The largest value $\eta$ = 1.071 is for the whole rectangular region presented in Figure 11a of which effective diameter is 205 Å.

Figure 13

Phason-space distributions of other two domains observed in Figure 10. Domains shown in (d) and (e) correspond to those surrounded by (orange) lines in Figure 9. The former ranges between $240 < x < 350$ and $30 < y < 170$, and the latter between $170 < x < 310$ and $200 < y < 340$. Their distributions in the phason space are presented in (b) and (c), respectively, using the same color scale as in Figure 10. Basis vectors $e_{1\perp}$- $e_{4\perp}$ are shown in (a).

Figure 14

Structure model of the dodecagonal quasicrystal corresponding to the fourth largest circle in Figure 11a. While sites with fractional occupancy appeared near the surface by simple application of the decoration scheme in Tables II and III, they are now fully occupied. Open (blue) and full (green) circles denote sites occupied by "M" and "M/Si "atoms, respectively. Six arrows indicate direction of quasi-lattice lines. An example of quasi-lattice line is drawn.

Figure 15

Calculated X-ray diffraction patterns of the structure model presented in Figure 14. (a) Plane at $h_5 = 0$, (b) $h_5 = 1$, and (c) $h_5 = 2$. Notice systematic extinction in (b).

Figure 16

Powder X-ray diffraction patterns of the dodecagonal quasicrystal. (a) Calculated pattern for the structure model shown in Figure 14. Indices of reflections are inserted.
(b) Experimentally measured intensity distribution. Some small artifacts are due to subtraction process of coexisting β-Mn type phase. See [24] for details.

Table I  Statistics of geometrical units and configurations examined in experimental and theoretical tilings. Types of the triangles and squares are assigned by two vectors starting from the same origin (see Figure 1b). Two types of the vertices $3^6$ are assigned by three edges of hexagons. The values for the theoretical models were calculated in the finite models presented in Figure 2.

\* Local configuration cannot be determined at the outermost region.

\*\* $3^6$ vertex is not always the center of the dodecagon.

| Tiling | Fig. 8 Exp. | Fig. 14 Exp. | Fig. 2(a) Deterministic | Fig. 2(b) Random |
|---|---|---|---|---|
| Number of tiles | 7540 | 1090 | 1180 | 1225 |
| Number of triangles | 5273 | 764 | 830 | 861 |
| Triangle 1  ($a_{1//}$, $a_{1//} + a_{3//}$) | 1238 | 180 | 199 | 218 |
| Triangle 2  ($a_{1//} + a_{3//}$, $a_{3//}$) | 1241 | 179 | 207 | 218 |
| Triangle 3  ($a_{2//}$, $a_{2//} + a_{4//}$) | 1397 | 202 | 210 | 212 |
| Triangle 4  ($a_{2//} + a_{4//}$, $a_{4//}$) | 1397 | 203 | 214 | 212 |
| Number of squares | 2262 | 325 | 350 | 363 |
| Square 1  ($a_{1//}$, $a_{2//} + a_{4//}$) | 816 | 111 | 118 | 121 |
| Square 2  ($a_{2//}$, $a_{3//}$) | 710 | 115 | 116 | 121 |
| Square 3  ($-a_{4//}$, $a_{1//} + a_{3//}$) | 736 | 99 | 116 | 121 |
| Number of 30° rhombi | 5 | 1 | 0 | 0 |
| Number ratio: $N_t/N_s$ | 2.331 | 2.351 | 2.371 | 2.372 |
| Number of vertices\* | 5093 | 760 | 830 | 857 |
| $3^6$ vertices\*\* | 341 | 47 | 52 | 63 |
| Full circle ($a_{2//}$, $a_{2//} + a_{4//}$, $a_{4//}$) | 214 | 31 | 32 | 33 |
| Open circle ($a_{1//}$, $a_{1//} + a_{3//}$, $a_{3//}$) | 127 | 16 | 20 | 30 |
| $3^2 \cdot 4 \cdot 3 \cdot 4$ vertices | 4231 | 599 | 632 | 475 |
| $3^3 \cdot 4^2$ vertices | 189 | 20 | 42 | 200 |
| $4^4$ vertices | 1 | 0 | 0 | 12 |
| Number of dodecagons\*\* | 335 | 47 | 52 | 11 |
| Full circle ($a_{2//}$, $a_{2//} + a_{4//}$, $a_{4//}$) | 208 | 31 | 32 | 7 |
| Open circle ($a_{1//}$, $a_{1//} + a_{3//}$, $a_{3//}$) | 127 | 16 | 20 | 4 |
| $\eta$ | 1.960 | 0.903 | 0.780 | 1.072 |

Table II  Atomic decoration of square. Coordinates *x*, *y* and *z* in orthogonal system with unit lengths of *a* and *c*, coordination number, occupation, atom types, and number of atoms in a cell are listed. Origin of the coordinate is set at the center of the square. See Figure 3a for the sites A - F. This square includes 8 atoms in total. For "M" and "M/Si", see text.

| Site | x   | y   | z    | C.N. | Occu. | Type | Number          |
|------|-----|-----|------|------|-------|------|-----------------|
| A    | 1/2 | 1/2 | 1/4  | 14   | 1/4   | M    | 1/4 × 4 = 1     |
| B    | 1/2 | 1/2 | -1/4 | 14   | 1/4   | M    | 1/4 × 4 = 1     |
| C    | 0   | 1/2 | 0    | 12   | 1/2   | M/Si | 1/2 × 2 = 1     |
| D    | 1/2 | 0   | 1/2  | 12   | 1/2   | M/Si | 1/2 × 2 = 1     |
| E    | 1/4 | 0   | 0    | 14   | 1     | M    | 1 × 2 = 2       |
| F    | 0   | 1/4 | 1/2  | 14   | 1     | M    | 1 × 2 = 2       |

Table III   Atomic decoration of equilateral triangle. Origin is set at the center of the triangle. Coordinates *x*, *y* and *z* are not in hexagonal, but in orthogonal system. See Figure 3d for the sites G - J. A triangle includes 3.5 atoms. For "M" and "M/Si", see text.

| Site | x   | y         | z   | C.N. | Occu. | Type | Number           |
|------|-----|-----------|-----|------|-------|------|------------------|
| G    | 1/2 | $-\sqrt{3}/6$ | 1/4 | 14   | 1/6   | M    | $1/6 \times 3 = 1/2$ |
| H    | 1/2 | $-\sqrt{3}/6$ | -1/4 | 14  | 1/6   | M    | $1/6 \times 3 = 1/2$ |
| I    | 1/4 | $\sqrt{3}/12$ | 0   | 12   | 1/2   | M/Si | $1/2 \times 3 = 3/2$ |
| J    | 0   | 0         | 1/2 | 15   | 1     | M    | 1                |

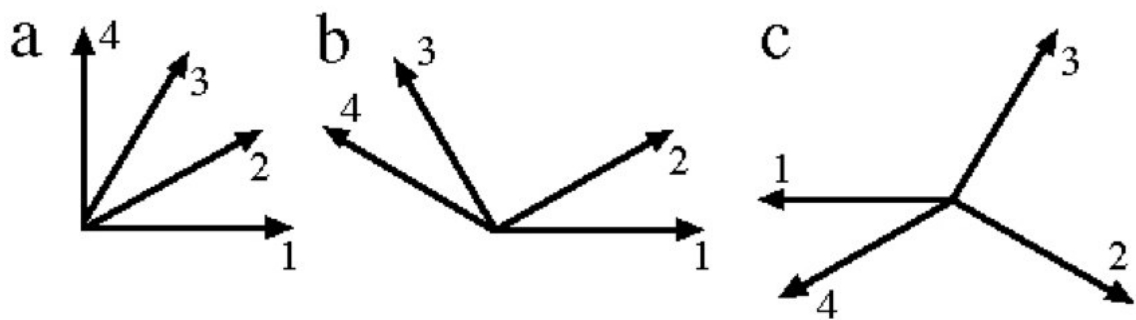

Fig. 1

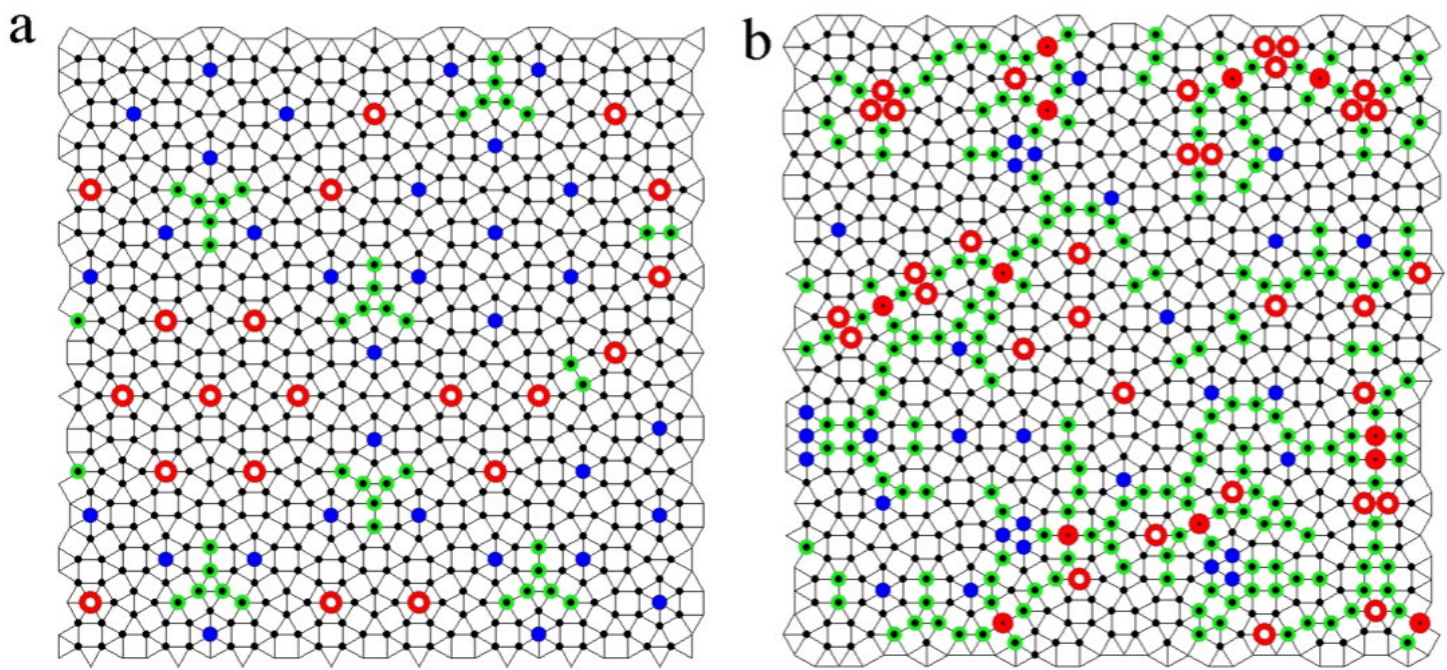

Fig. 2

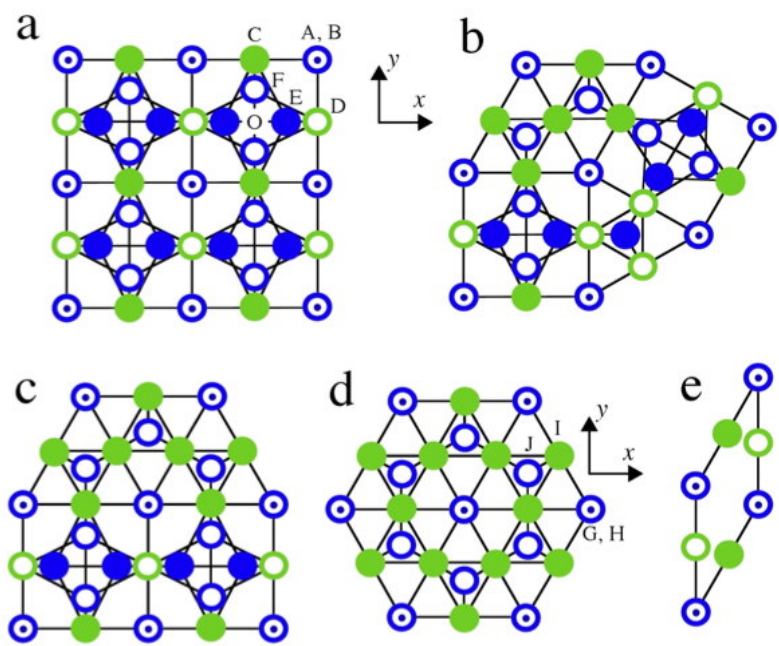

Fig. 3

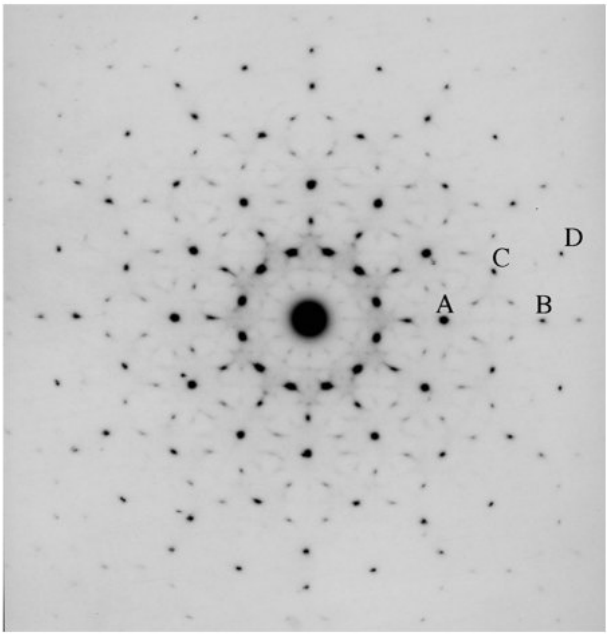

Fig. 4

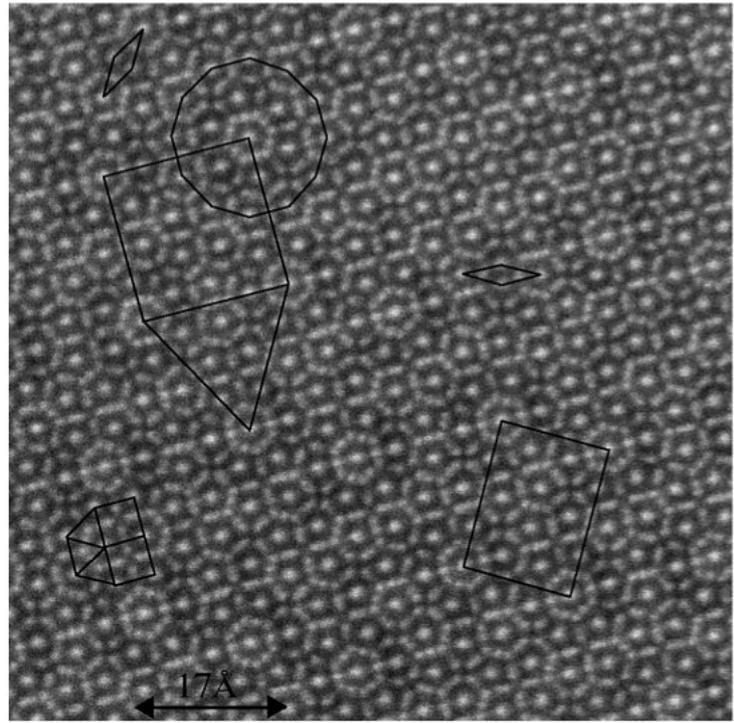

Fig. 5

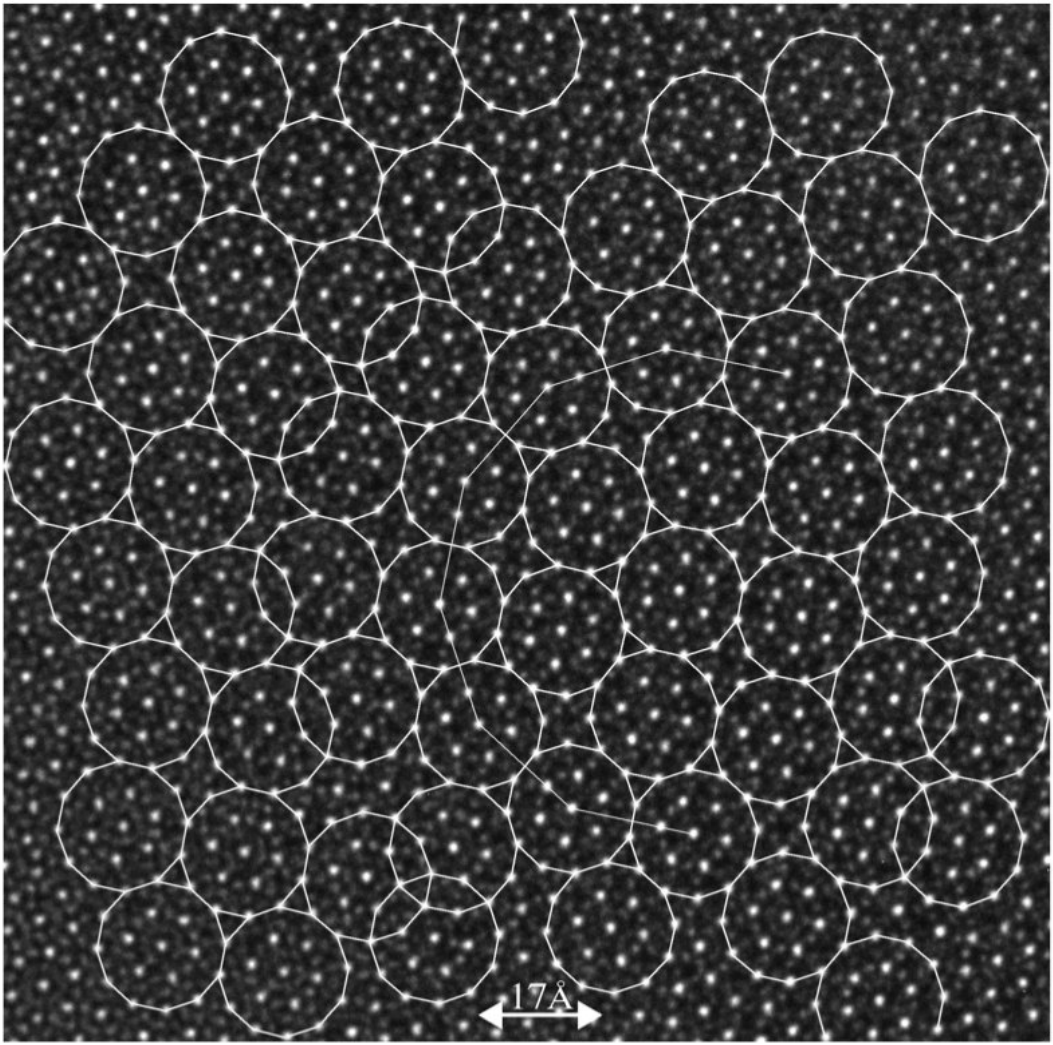

Fig. 6

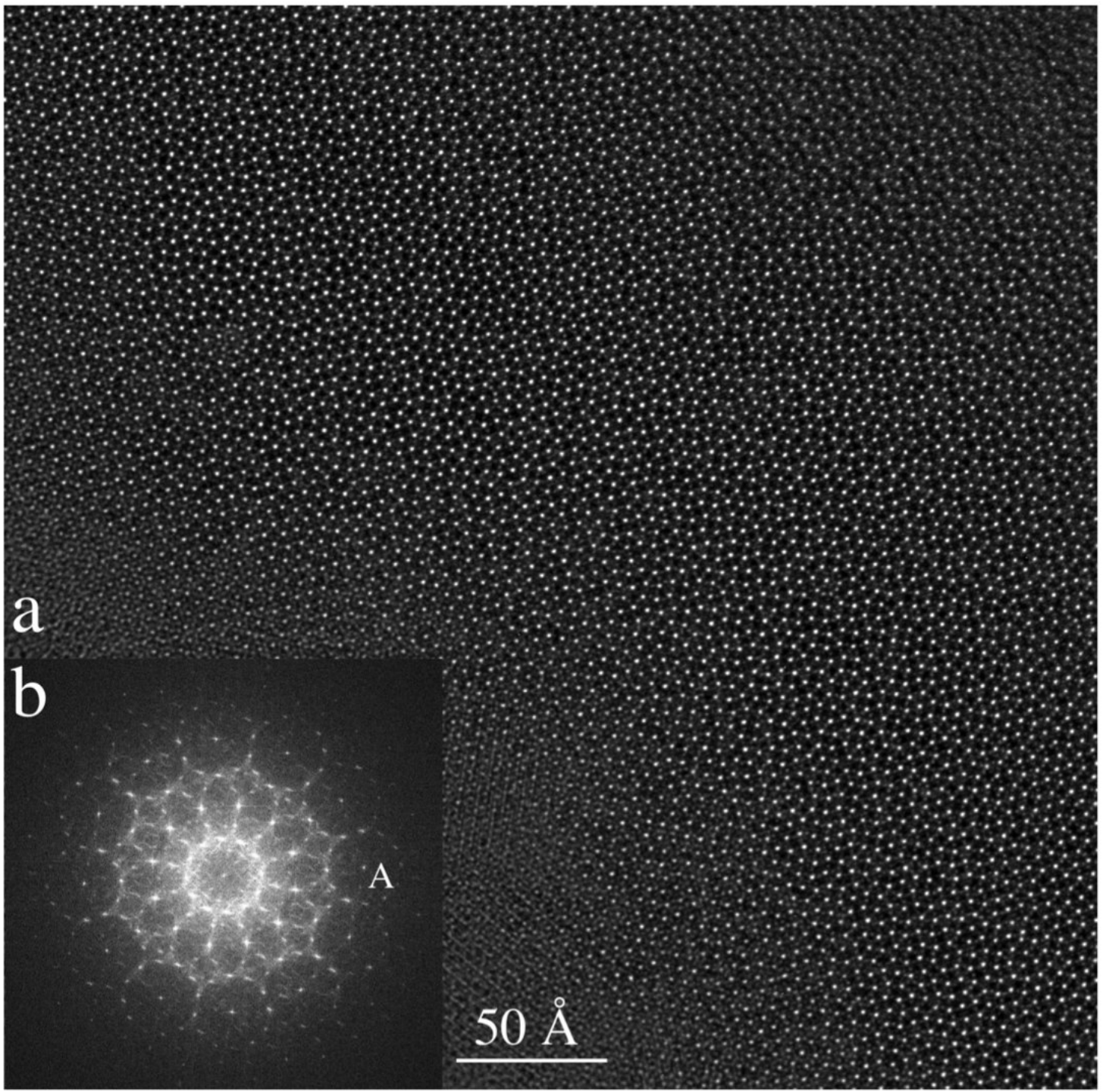

Fig. 7

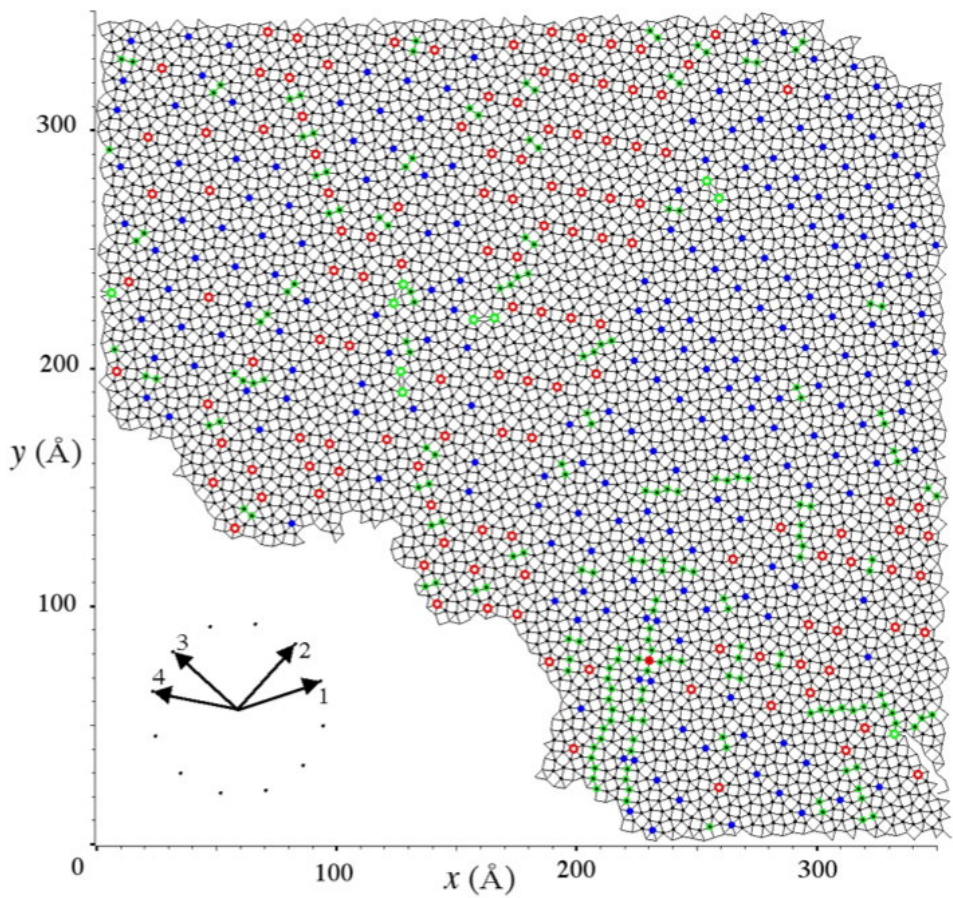

Fig. 8

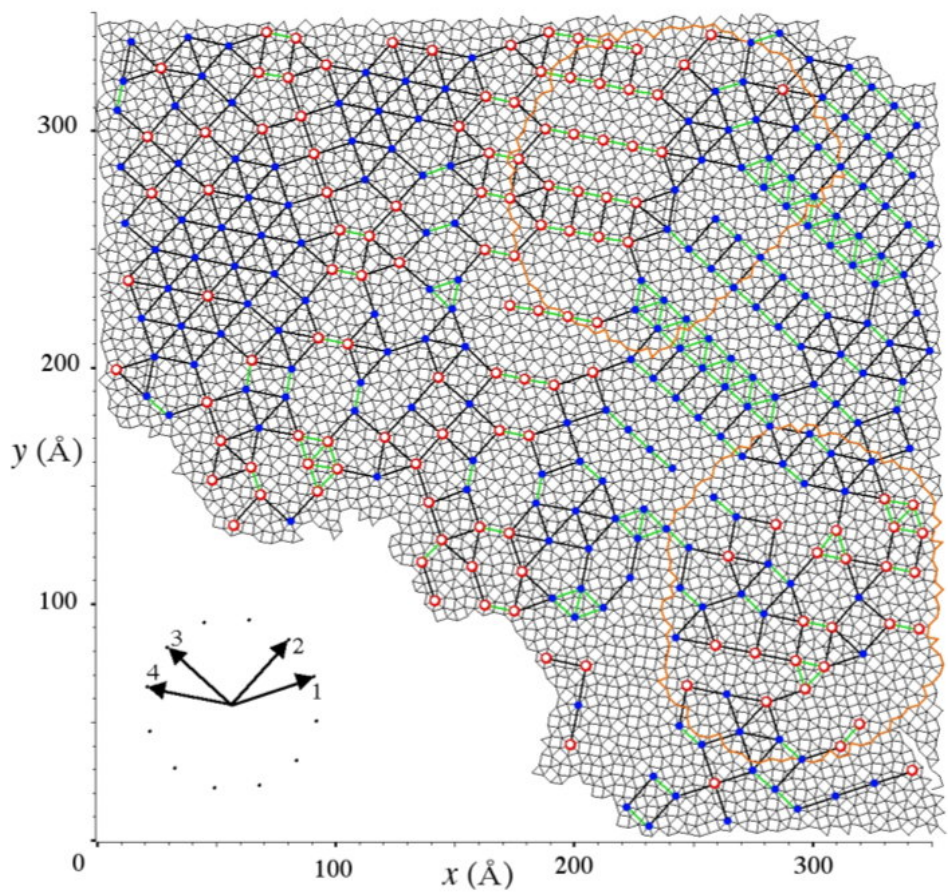

Fig. 9

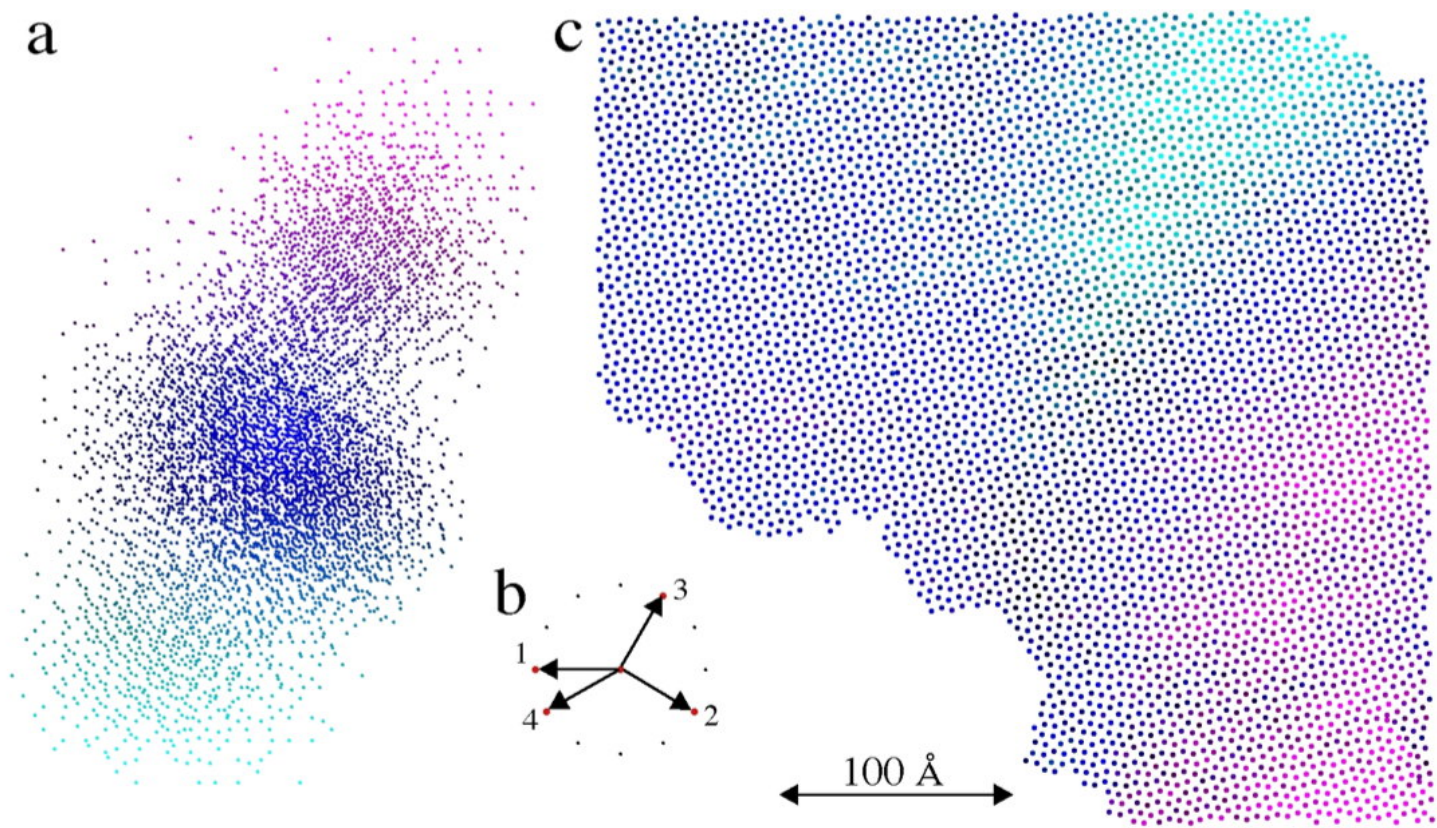

Fig. 10

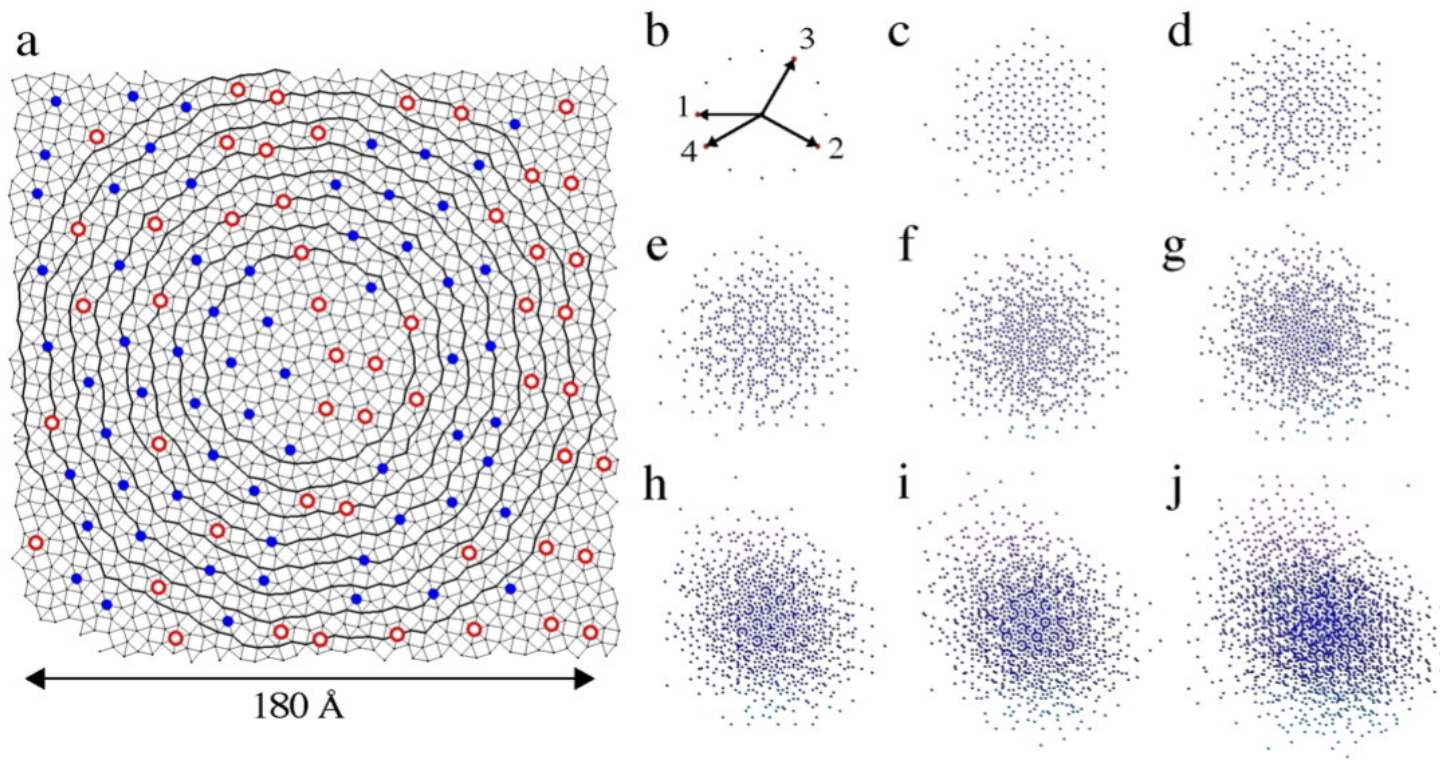

Fig. 11

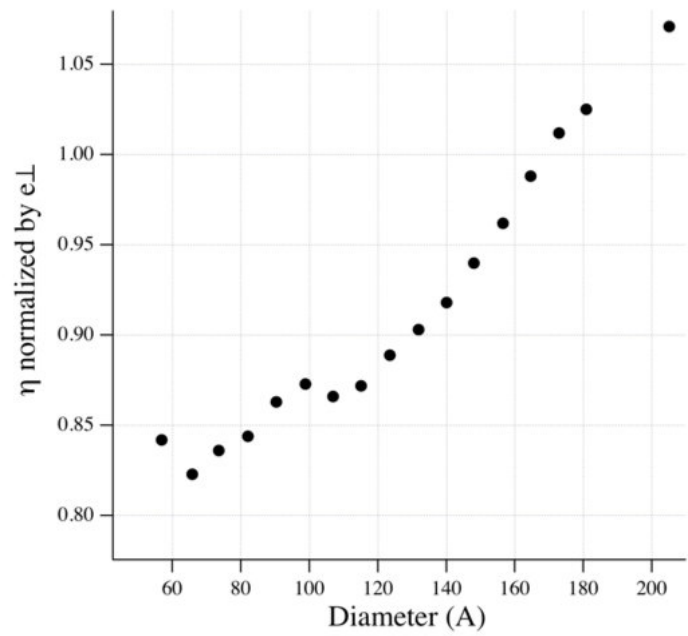

Fig. 12

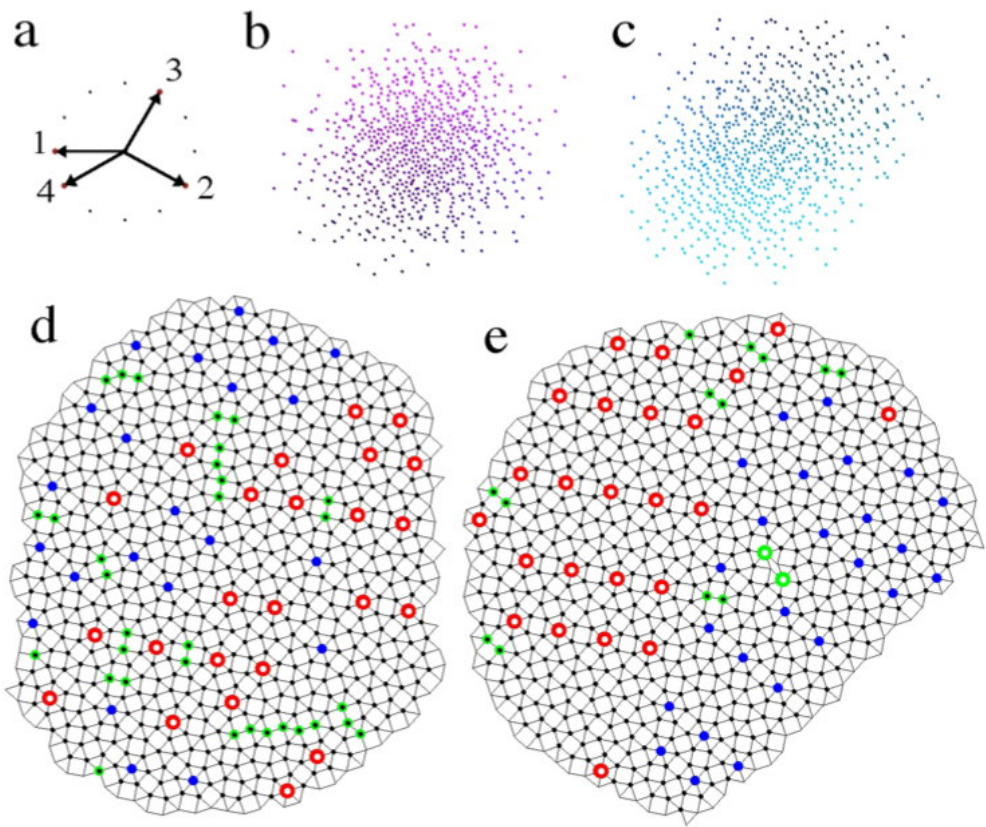

Fig. 13

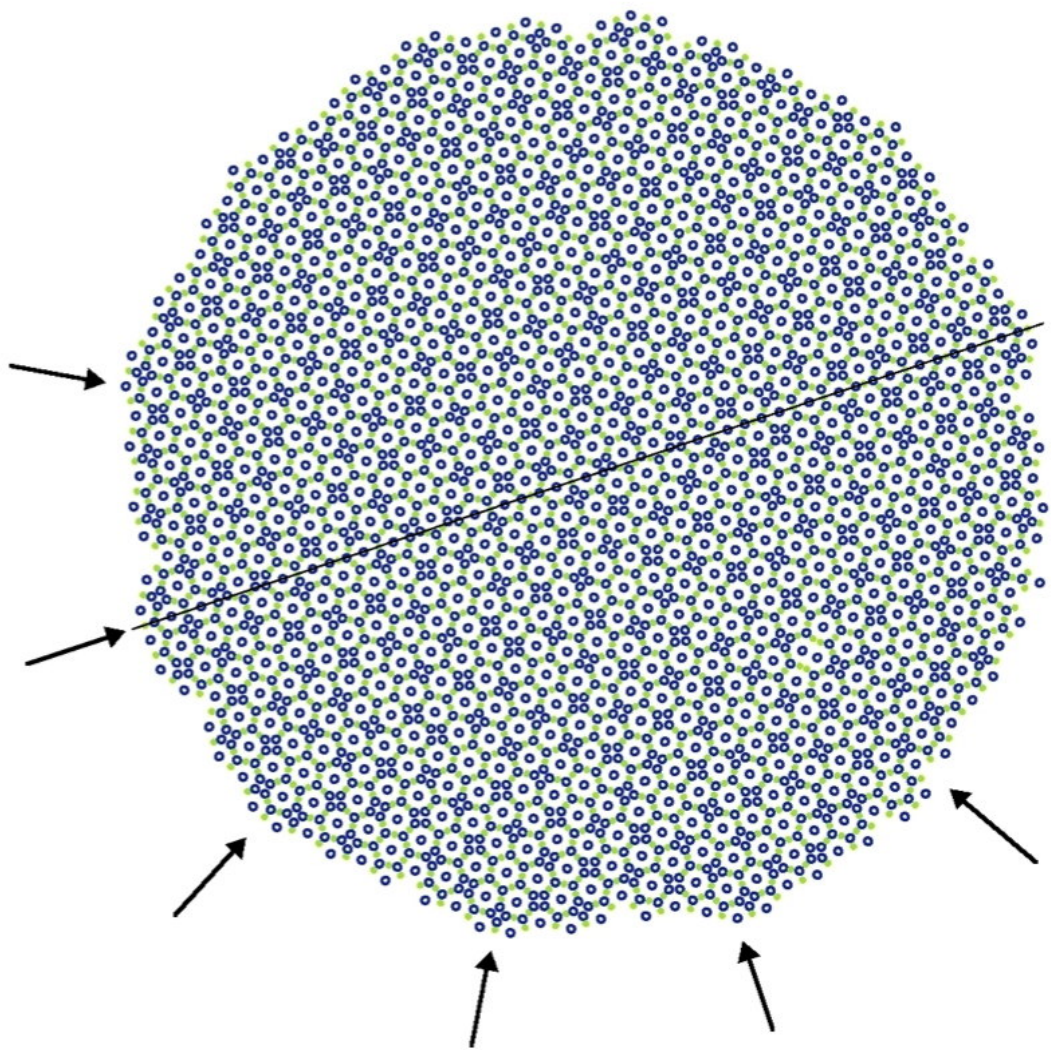

Fig. 14

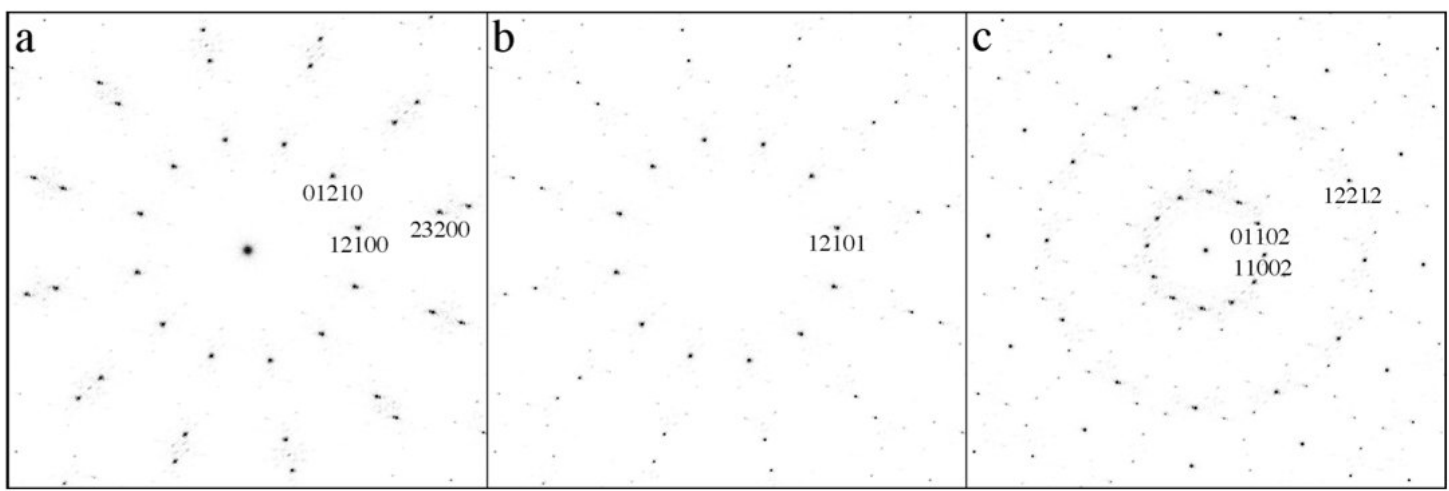

Fig. 15

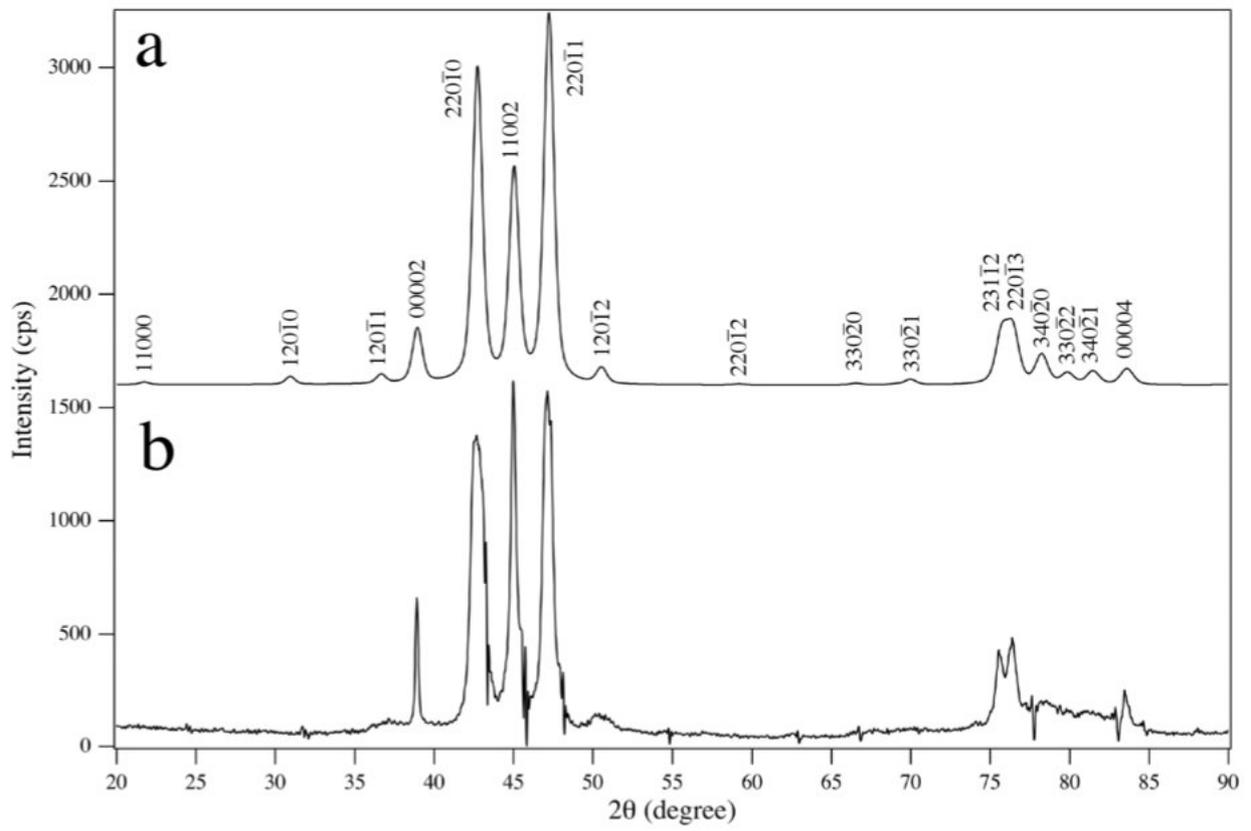

Fig. 16